\documentclass[11pt]{article}

\usepackage[margin=.9in]{geometry}
\usepackage[numbers]{natbib}
\usepackage{amsfonts,amsmath,amssymb,amsthm}
\usepackage{graphicx}
\usepackage[colorlinks,citecolor=green,linkcolor=blue,bookmarks=false,hypertexnames=true]{hyperref}

\newtheorem{definition}{Definition}
\newtheorem{remark}{Remark}

\renewcommand{\d}[1]{\ensuremath{\operatorname{d}\!{#1}}}
\newcommand{\D}[1]{\ensuremath{\operatorname{D}\!{#1}}}

\def\x{\mathbf x}
\def\v{\mathbf v}
\def\va{\mathbf v_\text a}
\def\vf{\mathbf v_\text f}
\def\vp{\mathbf v_\text p}
\def\u{\mathbf u}

\def\F{\mathbf F}

\begin{document}

\title{Dynamics of inertial particles on the ocean surface with unrestricted reserve buoyancy}

\author{F.\ J.\ Beron-Vera\\ Department of Atmospheric Sciences\\
Rosenstiel School of Marine, Atmospheric \& Earth Science\\ University
of Miami\\ Miami, Florida, USA\\fberon@miami.edu}

\date{Started: March 2, 2019; this version: \today.}%

\maketitle

\begin{abstract}
  The purpose of this note is to present an enhancement to a Maxey--Riley theory proposed in recent years for the dynamics of inertial particles on the ocean surface. This model upgrade removes constraints on the \emph{reserve buoyancy}, defined as the fraction of the particle volume above the ocean surface. The refinement results in an equation that correctly describes both the neutrally buoyant and fully buoyant particle scenarios.
\end{abstract}

\section{Introduction}

In \cite{Beron-etal-19-PoF}, a Maxey--Riley \cite{Maxey-Riley-83} framework was presented to describe the dynamics of floating finite-size or \emph{inertial} particles on the ocean surface, hereafter called \emph{BOM model}, expanding on a model for submerged particles in the ocean \cite{Beron-etal-15}. The fluid mechanics Maxey--Riley equation is a Newton's second law that accounts for various forces acting on spherical inertial particles carried by fluid flow \cite{Cartwright-etal-10}. The BOM theory has been validated both under natural field conditions \cite{Olascoaga-etal-20, Miron-etal-20-GRL} and under controlled lab settings \cite{Miron-etal-21-Chaos}, and a previous model \cite{Beron-etal-16} successfully explained garbage patches in the subtropical gyres of the ocean. Moreover, the BOM model has been extended to simulate networks of elastically interacting floating inertial particles \cite{Beron-Miron-20}, offering a simplified model for \emph{Sargassum} seaweed transport. Several results pertaining to the BOM model and its extensions are reviewed in \cite{Beron-21-ND}, additional laboratory experiments utilizing the BOM model are discussed in \cite{Olascoaga-etal-23}, and an application of the ``elastic'' BOM model to \emph{Sargassum} dynamics in the Caribbean Sea is detailed in \cite{Andrade-etal-22}.

The purpose of this short paper is to introduce an enhanced version of the BOM model that removes limitations on \emph{reserve buoyancy}, $\sigma$, which denotes the portion of the particle's volume that stays above the ocean surface (Sec.\@~2). This improvement is motivated by the observation made in \cite{Olascoaga-etal-20}, who pointed out that the water-to-particle density ratio cannot be too large---a limitation that we seek to relax. In doing this, the dependence on $\sigma$ of the projected length of the submerged particle piece, a factor in the drag force formula, is retrospectively determined using existing data on the windage parameter (Sec.\@~3). We complement the enhanced model with a reduced-order model that is asymptotically valid over the long term (Sec.\@~4). Finally, we consider the incorporation of general particle interactions and demonstrate how to account for the effects of Earth's curvature (Sec.\@~5).  Remarks have been interspersed in the text to clarify ideas and correct propagated typographical errors in earlier works.

\section{Setup}

Let $\x = (x, y)$ denote the position in a domain $D$ of the $\beta$-plane, where $x$ (eastward) and $y$ (northward) represent its coordinates. The domain $D \subset \mathbb{R}^2$ rotates with an angular velocity of $\frac{1}{2}f$, where $f = f_0 + \beta y$ denotes the Coriolis parameter. Let $z$ be the vertical coordinate, and let $t$ represent time. Imagine two homogeneous fluid layers stacked on top of each other, separated by an interface fixed at $z = 0$. The bottom layer, which represents water, has a density $\rho$, while the upper lighter layer represents air and has a density $\rho_{\mathrm{a}}$. Denote the dynamic viscosities of water and air as $\mu$ and $\mu_{\mathrm{a}}$, respectively. The velocities of water and air near the interface are represented as $\v(\x, t)$ and $\va(\x, t)$ accordingly. Finally, consider a solid spherical particle with radius $a$, presumed sufficiently small, and density $\rho_{\mathrm{p}}$ floating at the air--sea interface.

\begin{definition}[Reserve volume]
  Let
  \begin{equation}
    \delta : = \frac{\rho}{\rho_\mathrm{p}},\quad 
    \delta_\mathrm{a} : = \frac{\rho_\mathrm{a}}{\rho_\mathrm{p}}
    \equiv \delta\frac{\rho_\mathrm{a}}{\rho}.
    \label{eq:dta}
  \end{equation}
  Let $0 \le \sigma \le 1$ be the \textbf{reserve buoyancy}, that is, the fraction of emergent (above the ocean surface) particle volume. Static vertical stability of the particle (Archimedes' principle),
  \begin{equation}
    (1-\sigma)\delta + \sigma\delta_\mathrm{a} = 1,
    \label{eq:arch}
  \end{equation}
  is satisfied for
  \begin{equation}
    \sigma = \frac{\delta-1}{\delta-\delta_\mathrm{a}} \equiv \frac{1-\delta^{-1}}{1-\rho_\mathrm{a}/\rho},
    \label{eq:sigma}
  \end{equation}
  which requires
  \begin{equation}
    1\le \delta \le \frac{\rho}{\rho_\mathrm{a}},\quad
    \frac{\rho_\mathrm{a}}{\rho} \le \delta_\mathrm{a} \le 1.
  \end{equation}
\end{definition}

The configurations for \emph{neutrally buoyant} and \emph{maximally buoyant} particles reside at opposite extremes. The neutrally buoyant condition is characterized by $\sigma = 0$, which translates to $\delta = 1$ and $\delta_\mathrm{a} = 0$. Conversely, the maximally buoyant case is characterized by $\sigma = 1$, formally achieved by making $\delta = \rho/\rho_\mathrm{a}$ and $\delta_\mathrm{a} = 1$.

\begin{figure}[t!]
  \centering%
  \includegraphics[width=\textwidth]{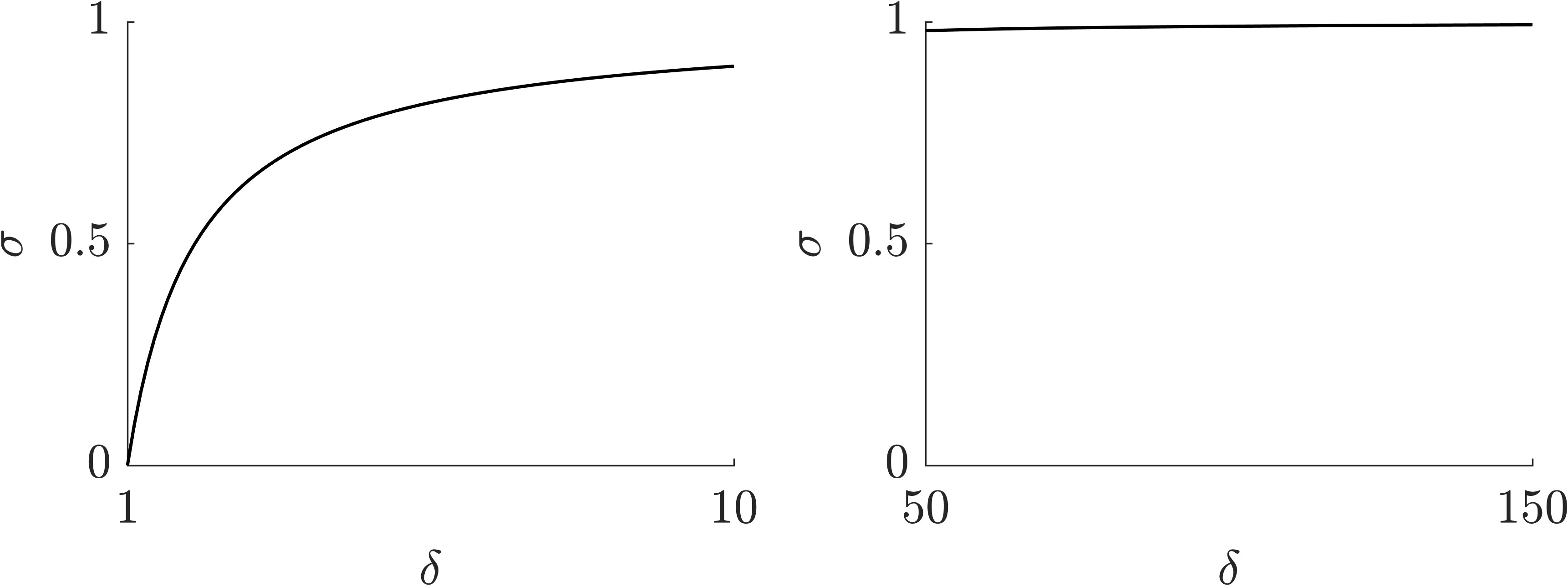}%
  \caption{Reserve buoyancy as a function of the particle-to-water density ratio, which, unlike in the BOM theory, can be taken as large as desired.}
  \label{fig:sigma}%
\end{figure}

\begin{remark}
Two points should be highlighted:
\begin{enumerate}
    \item Given that $\rho_\mathrm{a}/\rho$ is significantly smaller than 1, the reserve buoyancy $\sigma$ can be approximated by $1 - \delta^{-1}$ (Fig.\@~\ref{fig:sigma}). Observe the distinct behaviors of $\sigma$ and $\delta$, the preferred parameter in the BOM theory. When $\delta$ equals 1, $\sigma$ vanishes, representing the neutrally buoyant configuration. In contrast, as $\delta$ increases, $\sigma$ approaches 1, corresponding to the maximally buoyant configuration. This cannot be described by the BOM theory as it imposes a constraint on $\delta$, preventing it from becoming too large \textup{\cite{Olascoaga-etal-20}}.
    
    \item The value of the parameter $\delta_\mathrm{a}$ does not have to be small in all cases. It is close to 0 when $\delta$ is finite (meaning $\sigma$ remains below 1), but it nears 1 as $\delta$ increases significantly (or in other words, as $\sigma$ approaches 1).
\end{enumerate}
\end{remark}

\begin{figure}[t!]
  \centering%
  \includegraphics[width=.75\textwidth]{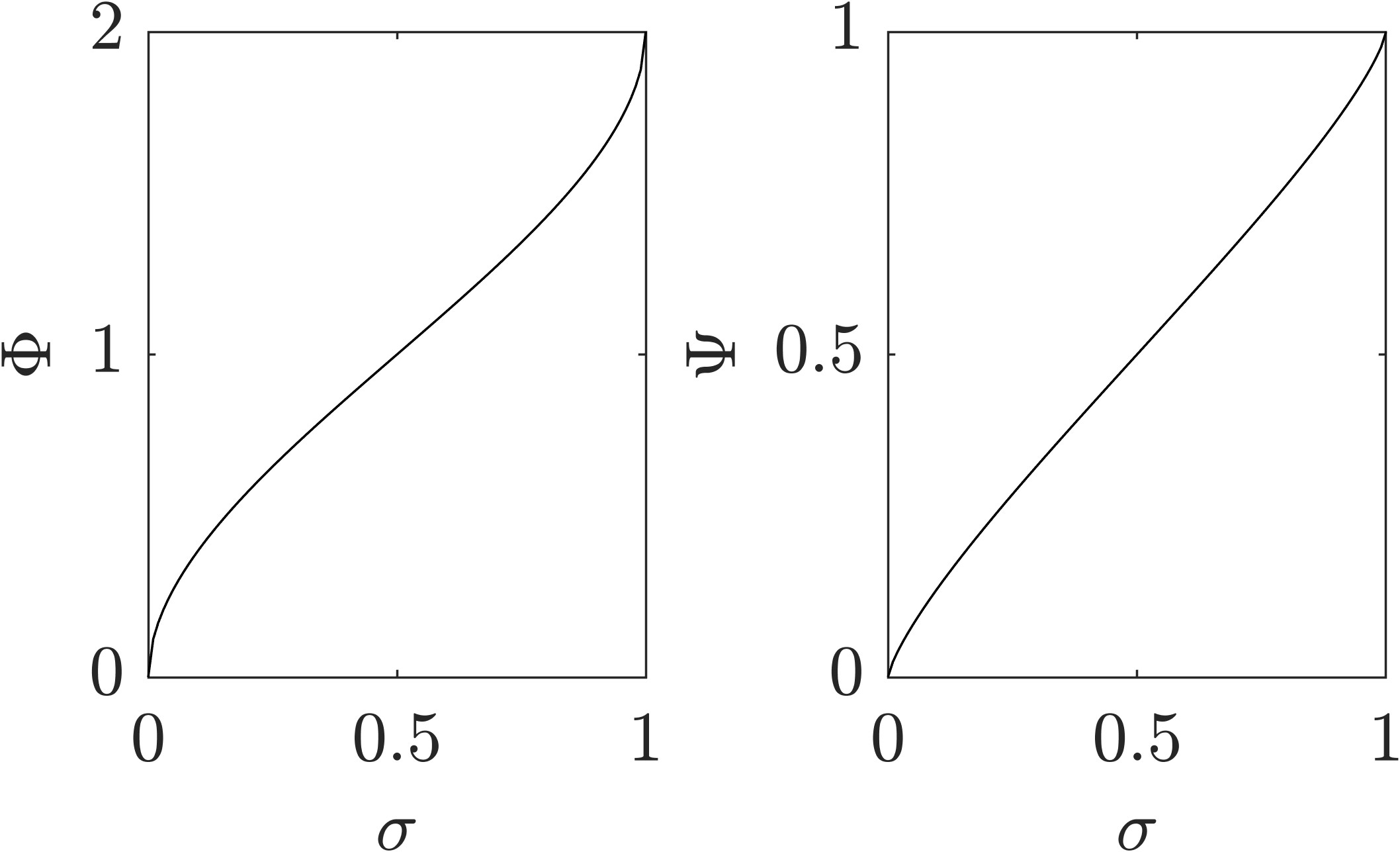}%
  \caption{As a function of reserve buoyancy, height of the emerged spherical cap normalized by radius (left) and area of the of the emerged spherical cap normalized by maximal sectional area (right).}
  \label{fig:PhiPsi}%
\end{figure}

The height ($h_\mathrm{a}$) of the emerged spherical cap can be described using $\sigma$. This results from matching its volume formula, expressed with $a$ and $h_\mathrm{a}$, to the volume of the emerged spherical cap. Specifically,
\begin{equation}
  \frac{\pi h_\mathrm{a}^2}{3}(3a - h_\mathrm{a}) =
  \sigma\frac{4}{3}\pi a^3,
  \label{eq:cubic}
\end{equation}
whose only physically meaningful root is
\begin{equation}
  h_\mathrm{a}/a = \Phi := \frac{\mathrm{i}\sqrt{3}}{2}
  \left(\frac{1}{\varphi}-\varphi\right) - \frac{1}{2\varphi} -
  \frac{\varphi}{2} + 1
  \label{eq:Phi}
\end{equation}
where
\begin{equation}
  \varphi := \sqrt[3]{\mathrm{i}\sqrt{1-(1 - 2\sigma)^2} +
  1 - 2\sigma}
  \label{eq:phi}
\end{equation}
(Fig.\@~\ref{fig:PhiPsi}, left panel). The depth ($h$) of the submerged spherical cap,
\begin{equation}
  h = (2-\Phi)a.
\end{equation}

To be referenced in the future, the area of the emerged spherical cap projected in the direction of the flow, $A_\mathrm{a}$, is clearly given by
\begin{equation}
  A_\mathrm{a} = \pi\Psi a^2,\quad 
  \Psi := \pi^{-1}\cos^{-1}(1-\Phi)
  - \pi^{-1}(1-\Phi) \sqrt{1-(1-\Phi)^2},
  \label{eq:Aa}
\end{equation}
a function of $\sigma$ exclusively (Fig.\@~\ref{fig:PhiPsi}, right panel). In turn, the immersed projected area, denoted $A$, is equal to
\begin{equation}
  A = \pi a^2 - A_\mathrm{a} = \pi(1 - \Psi)a^2.
  \label{eq:A}
\end{equation}

\section{Maxey--Riley equation for unrestricted reserve buoyancy}

Fluid variables and parameters, denoted with a subscript f, differ for water and air, for example,
\begin{equation}
  \vf(\x,z,t) = 
  \begin{cases} 
    \va(\x,t) & \text{if } z \in (0,h_\mathrm{a}],\\
    \v(\x,t)  & \text{if } z \in [-h,0).
  \end{cases}
\end{equation}
Following \cite{Beron-etal-19-PoF}, we write
\begin{equation}
  \dot\v_\text{p} + f \vp^\perp = \langle \F_\mathrm{flow}\rangle + \langle \F_\mathrm{mass}\rangle + \langle \F_\mathrm{lift}\rangle + \langle \F_\mathrm{drag}\rangle,
  \label{eq:mr}
\end{equation}
where $\perp$ denotes $+\frac{\pi}{2}$-rotation and $\langle\,\rangle$ is an average over $z\in [-h,h_\mathrm{a}]$. Here, $\vp(t)$ represents the particle velocity, $\F_\text{flow}$ is the force per unit density of the undisturbed flow, $\F_\text{mass}$ is the added mass force, $\F_\text{lift}$ is the lift force from a horizontally sheared flow, and $\F_\text{drag}$ is the drag force from fluid viscosity. All forcing terms in Eq.\@~\eqref{eq:mr} are part of the Maxey--Riley equation \cite{Maxey-Riley-83}. Exceptions are the Coriolis force term \cite{Beron-etal-15} and the lift force term \cite{Montabone-02}. For details, see \cite{Beron-etal-19-PoF, Beron-21-ND}.

In contrast to \cite{Beron-etal-19-PoF}, we calculate the vertical averages in Eq.\@~\eqref{eq:mr} \emph{retaining all terms irrespective of their insignificance relative to the air-to-particle density ratio ($\delta_\mathrm{a}$) when considering a finite water-to-particle density ratio ($\delta$)}. We begin with the flow force:
\begin{align}
  \langle \F_\mathrm{flow}\rangle 
  = {} &
  \frac{1}{2a}\int_{-h}^{h_\mathrm{a}}
  \frac{m_\mathrm{f}}{m_\mathrm{p}}\left(\frac{\D{\vf}}{\D{t}} + f
  \vf^\perp\right)\d{z}\nonumber\\
  = {} &
  \frac{1}{2a}\int_{(\Phi-2)a}^{0}
  \frac{\sigma\frac{4}{3}\pi a^3\rho}{\frac{4}{3}\pi 
  a^3\rho_\mathrm{p}}\left(\frac{\D{\v}}{\D{t}} + f
  \v^\perp\right)\d{z}\nonumber\\
  & +
  \frac{1}{2a}\int_{0}^{\Phi a}
  \frac{\sigma\frac{4}{3}\pi a^3\rho_\mathrm{a}}{\frac{4}{3}\pi 
  a^3\rho_\mathrm{p}}\left(\frac{\D{\va}}{\D{t}} + 
  f \va^\perp\right)\d{z}\nonumber\\
  = {} & 
  \left(1-\frac{\Phi}{2}\right)(1-\sigma)\delta\left(\frac{\D{\v}}{\D{t}}
  + f \v^\perp\right) + 
  \frac{\Phi}{2}\sigma\delta_\mathrm{a}\left(\frac{\D{\va}}{\D{t}} + 
  f \va^\perp\right),
  \label{eq:FFavg}
\end{align}
where
\begin{equation}
  \frac{\D{\vf}}{\D{t}} = \partial_t \vf + v^x_\mathrm{f}\partial_x\vf + v^y_\mathrm{f}\partial_y\vf 
  \label{eq:dvdt}
\end{equation} 
is the material derivative of $\vf$ along a fluid trajectory, i.e., obtained by solving $\dot{\x} = \vf$.

Similarly, we compute
\begin{align}
  \langle \F_\mathrm{mass}\rangle &= \frac{1}{2a}\int_{-h}^{h_\mathrm{a}}
  \frac{\frac{1}{2}m_\mathrm{f}}{m_\mathrm{p}}\left(\frac{\D{\vf}}{\D{t}}
  + f \vf^\perp - \dot\v_\mathrm{p} - f\vp^\perp\right)\d{z}\nonumber\\ 
  &=
  \frac{1}{2}\left(1-\frac{\Phi}{2}\right)(1-\sigma)\delta\left(\frac{\D{\v}}{\D{t}}
  + f \v^\perp - \dot\v_\mathrm{p} - f \v_\mathrm{p}^\perp\right) +
  \frac{1}{2}\frac{\Phi}{2}\sigma\delta_\mathrm{a}\left(\frac{\D{\va}}{\D{t}}
  + f \va^\perp - \dot\v_\mathrm{p} - f \vp^\perp\right)
  \label{eq:AMavg}
\end{align}
and, letting
\begin{equation}
  \omega_\mathrm{f} = \partial_x v^y_\mathrm{f} - \partial_y v^x_\mathrm{f},
  \label{eq:vor}
\end{equation}
that is, the vorticity of the fluid, we likewise compute
\begin{align}
  \langle \F_\mathrm{lift}\rangle 
  &=
  \frac{1}{2a}\int_{-h}^{h_\mathrm{a}}
  \frac{\frac{1}{2}m_\mathrm{f}}{m_\mathrm{p}}\omega_\mathrm{f} 
  \left(\vf - \vp\right)^\perp\d{z}\nonumber\\
  &=
  \frac{1}{2}\left(1-\frac{\Phi}{2}\right)(1-\sigma)\delta\omega 
  \left(\v - \vp\right)^\perp +
  \frac{1}{2}\frac{\Phi}{2}\sigma\delta_\mathrm{a}\omega_\mathrm{a} 
  \left(\va - \vp\right)^\perp.
  \label{eq:FLavg}
\end{align}

To determine the drag force, it is necessary to select suitable characteristic projected length scales for the submerged and emergent sections of the particle, denoted by $\ell$ and $\ell_\mathrm{a}$, respectively. We reasonably assume that $\ell$ and $\ell_\mathrm{a}$ both are smaller or equal to $2a$, with $\ell_\mathrm{a} = 2a - \ell$ as a function of $\sigma$, constrained by $\ell(0) = 2a$ and $\ell(1) = 0$. Recognizing the heuristic nature of this approach, given the absence of an explicit formula for the drag force on spherical caps, the natural way forward is to leave $\ell(\sigma)$ for empirical determination from appropriate measurements. With this in mind, we compute:
\begin{align}
  \langle \F_\mathrm{drag}\rangle
  = & {} 
  \frac{1}{2a}\int_{-h}^{h_\mathrm{a}}
  \frac{12\mu_\mathrm{f}\frac{A_\mathrm{f}}{\ell_\mathrm{f}}}{m_\mathrm{p}}
  (\vf - \vp)\d{z}\nonumber\\
  = & {}
  \frac{1}{2a}\int_{(\Phi-2)a}^{0}
  \frac{12\mu\frac{\pi(1-\Psi)a^2}{\ell(\sigma)}}{\frac{4}{3}\pi a^3\rho_\mathrm{p}}
  (\v - \vp)\d{z}\nonumber\\
  & + 
  \frac{1}{2a}\int_0^{\Phi a}
  \frac{12\mu_\mathrm{a}\frac{\pi\Psi a^2}{2a - \ell(\sigma)}}{\frac{4}{3}\pi a^3\rho_\mathrm{p}}
  (\va - \vp)\d{z}\nonumber\\
  = & {} 
  \frac{3}{2}\left(\frac{2}{3} +
  \frac{(1-\sigma)\delta}{3}+\Big(\sigma\delta_\mathrm{a}
  - \sigma\delta\Big)\frac{\Phi}{6}\right)\frac{\u - \vp}{\tau},
  \label{eq:SDavg}
\end{align}
where
\begin{equation}
  \u := (1-\alpha)\v + \alpha \va, 
  \label{eq:u}
\end{equation}
and the parameters
\begin{align}
  \tau &:= \frac{\frac{2}{3} +
  \frac{(1-\sigma)\delta}{3}+\Big(\sigma\delta_\mathrm{a}
  - (1-\sigma)\delta\Big)\frac{\Phi}{6}}{3\left((2-\Phi)(1-\Psi) +
   \frac{\ell(\sigma)}{2a-\ell(\sigma)}\gamma\Phi\Psi\right)\delta
  }\cdot \frac{a\ell(\sigma)}{\mu/\rho},\label{eq:tau}\\ 
  \alpha &:= \frac{\gamma \Phi\Psi}{(2-\Phi)(1-\Psi)
  \frac{2a-\ell(\sigma)}{\ell(\sigma)}+
  \gamma \Phi\Psi},\label{eq:alpha}\\
  \gamma &:= \frac{\mu_\mathrm{a}}{\mu}.
\end{align}

Finally, plugging Eqs.\@~\eqref{eq:FFavg}--\eqref{eq:SDavg} into Eq.\@~\eqref{eq:mr} and after some algebraic manipulation, we obtain the following Maxey--Riley equation:
\begin{equation}
  \dot \v_\mathrm{p} + \left(f +
  \tfrac{1}{3}R\omega + \tfrac{1}{3}R_\mathrm{a}\omega_\mathrm{a}\right)\vp^\perp + \tau^{-1} \vp = R\frac{\D{\v}}{\D{t}} + R\left(f + \tfrac{1}{3}\omega\right)\v^\perp + R_\mathrm{a}\frac{\D{\va}}{\D{t}} + R_\mathrm{a}\left(f + \tfrac{1} {3}\omega_\mathrm{a}\right)\va^\perp + \tau^{-1}\u, 
  \label{eq:MR}
\end{equation}
where
\begin{align}
  R &:= \frac{\left(1 - \frac{\Phi}{2}\right)(1-\sigma)\delta}{\frac{2}{3} +
  \frac{(1-\sigma)\delta}{3}+\Big(\sigma\delta_\mathrm{a}
  - (1-\sigma)\delta\Big)\frac{\Phi}{6}},\\
  R_\mathrm{a} &:= \frac{\frac{\Phi}{2}\sigma\delta_\mathrm{a}}{\frac{2}{3} +
  \frac{(1-\sigma)\delta}{3}+\Big(\sigma\delta_\mathrm{a}
  - (1-\sigma)\delta\Big)\frac{\Phi}{6}}.
  \label{eq:R}
\end{align}

Typically, the parameter $\gamma$ is approximately 0.0167. The parameter $\tau$ is positive and small, because $a$ is assumed to be small. The parameter $\alpha$ increases from 0 to 1 as $\sigma$ increases from 0 to 1, which can be expected regardless of the specific form of the projected length scale of the submerged spherical cup $\ell(\sigma)$. We only need to assume that $\ell$ decreases from $2a$ to 0 as $\sigma$ increases from 0 to 1. Finally, the parameter $R$ decreases from 1 to 0 as $\sigma$ increases from 0 to 1, whereas $R_\mathrm{a}$ increases from 0 to 1.

\begin{remark}
  The parameter $\tau$ can be identified with \textbf{Stokes' time}, which represents the time required for a particle to respond to inertial effects. The parameter $\alpha$ can be called \textbf{windage parameter} because it quantifies the contribution of air velocity to the \textbf{carrying flow velocity} $\mathbf{u}$, given by the convex combination of water and air velocities in Eq.\@~\eqref{eq:u}.

  Interestingly, $\alpha$ measures windage similarly to the ``leeway parameter'' commonly used in the search-and-rescue literature \textup{\cite{Breivik-etal-13}}. However, unlike $\alpha$, the leeway parameter is based primarily on ad hoc considerations rather than derived from first principles.
\end{remark}

In order to complete the formulation of the Maxey--Riley model, as specified in Eq.\@~\eqref{eq:MR}, it is required to evaluate $\ell(a)$. An evaluation of $\ell(a)$ can be achieved by using results from controlled laboratory measurements of the windage parameter $\alpha$ within the range $0 \le \delta \lessapprox 4$, as documented by \cite{Miron-etal-20-PoF}. This $\delta$-range corresponds to $0 \le \sigma \lessapprox 0.75$, as it follows from Eq.\@~\eqref{eq:sigma} as a consequence of the smallness of $\rho_\mathrm{a}/\rho$ in the noted finite $\delta$-range. These empirical observations suggest that, within the inferred $\sigma$-range, 
\begin{equation}
  \alpha \approx \frac{\gamma \Psi}{1 + (\gamma - 1)\Psi}.
  \label{eq:alpha-bom}
\end{equation}
Consequently, 
\begin{equation}
  \ell \approx (2 - \Phi)a = h, 
  \label{eq:ell}
\end{equation}
indicating that, over the range $0 \le \sigma \lessapprox 0.75$, the appropriate characteristic projected length scale of the submerged spherical cap is simply its depth. While this might appear self-evident, the drag force formula applies to complete spheres rather than spherical caps. Consistent with Eq.\@~\eqref{eq:ell}, 
\begin{equation}
  \tau \approx \frac{\frac{2}{3} + \frac{(\sigma-1)\delta}{3} + \Big(\sigma\delta_\mathrm{a} - (1-\sigma)\delta\Big) \frac{\Phi}{6}}{3\left(1 + (\gamma-1)\Psi\right)\delta} \cdot \frac{a^2}{\mu/\rho} \approx \frac{1 - \frac{1}{6}\Phi}{3(1 + (\gamma -1)\Psi\delta} \cdot \frac{a^2}{\mu/\rho}, 
\end{equation}
where the last approximation holds because $\sigma \approx 1 - \delta^{-1}$ independent of the value of $\delta$.  The derived formula for $\tau$ differs by a factor of $\delta^{-3}$ from the one obtained through empirical fitting to observational data reported in \cite{Olascoaga-etal-20}. However, these observations dealt with objects of various geometries, not just spheres, which required applying the heuristic shape correction introduced in \cite{Ganser-93}. Extending the closure in Eq.\@~\eqref{eq:ell} beyond the interval $0 \le \sigma \lessapprox 0.75$ will require further measurements. It is important to note that the interval $0 \le \sigma \lessapprox 0.75$ encompasses a substantial portion of the possible range of reserve buoyancy values. Therefore, it appears that Eq.\@~\eqref{eq:ell} already offers an empirically determined projected length well-supported by data.

\begin{remark}
  In the Appendix of \textup{\cite{Miron-etal-20-PoF}}, the expression in Eq.\@~\eqref{eq:alpha-bom} contains a typo where $1 - \gamma$ is used instead of $\gamma - 1$. This typographical error has been propagated to \textup{\citep{Miron-etal-20-GRL, Beron-Miron-20}}, and we wish to bring attention to this issue. 
\end{remark}

Lastly, all terms related to air in the Maxey--Riley model, Eq.\@~\eqref{eq:MR}, can be safely ignored when $\delta$ is finite (i.e., $\sigma$ is sufficiently smaller than 1), in which case $R_\mathrm{a}$ in Eq.\@~\eqref{eq:R} is small due to the smallness of $\delta_\mathrm{a}$. These air-related terms are excluded from the BOM equation. \emph{However, these terms will gain significance as $\delta$ becomes sufficiently large (or equivalently as $\sigma$ approaches 1), thereby expanding the applicability of the BOM model.}

\section{Relevant limits and slow manifold reduction}

For neutrally buoyant particles, i.e., when $\sigma = 0$, which corresponds to $\delta = 1$ and $\delta_\mathrm{a} = \rho_\mathrm{a}/\rho$, Eq.\@~\eqref{eq:MR} reduces to
\begin{equation}
  \dot\v_\mathrm{p} + \left(f + \tfrac{1}{3}\omega\right)\vp^\perp + \tau^{-1}\vp = \frac{\D{\v}}{\D{t}} + \left(f + \tfrac{1}{3}\omega\right)\v^\perp + \tau^{-1}\v, 
  \label{eq:MR-neutral} 
\end{equation}
with 
\begin{equation}
  \tau = \frac{a\ell(0)}{6\mu/\rho} = \frac{a^2}{3\mu/\rho}.  
  \label{eq:tau-neutral}
\end{equation} 
Conversely, in the case of maximally buoyant particles, i.e., for $\sigma = 1$, which translates to $\delta = \rho/\rho_\mathrm{a}$ and $\delta_\mathrm{a} = 1$, Eq.\@~\eqref{eq:MR}
simplifies to
\begin{equation}
  \dot\v_\mathrm{p} + \left(f + \tfrac{1}{3}\omega_\mathrm{a}\right)\vp^\perp + \tau^{-1} \vp = \frac{\D{\va}}{\D{t}} + \left(f + \tfrac{1}{3}\omega_\mathrm{a}\right)\va^\perp + \tau^{-1}\va, 
  \label{eq:MR-max} 
\end{equation}
with
\begin{equation}
  \tau = \frac{a\ell_\mathrm{a}(1)}{6\mu_\mathrm{a}/\rho_\mathrm{a}} = \frac{a^2}{3\mu_\mathrm{a}/\rho_\mathrm{a}}.  
  \label{eq:tau-max}
\end{equation}
The above equations exactly match the Maxey--Riley equation in fluid mechanics \cite{Maxey-Riley-83} when considering particles that have the same density as the carrying fluid (with the assumptions of motion on a $\beta$-plane, and the inclusion of lift force).

The movement of neutrally buoyant particles generally differs from the movement of water particles. Under specific conditions, it can be demonstrated to be unstable \cite{Haller-Sapsis-10}, irrespective of the Coriolis and lift effects \cite{Beron-etal-19-PoF, Beron-21-ND}. This applies mutatis mutandis to the motion of maximally buoyant particles with respect to that of air particles. These results are relevant in the short run as, over time, the movement of both neutrally buoyant and maximally buoyant particles aligns with the Lagrangian movement, meaning they align with the movement of water and air particles, respectively. 

\begin{remark}
  The motion of fully exposed particles is significantly more restricted by the assumption of a flat air--sea interface compared to the movement of partially or completely submerged particles. For these submerged particles, the influence of a wavy interface can be considered by incorporating a representation of the Stokes drift in the near-surface water velocity. This does not have an equivalent for the near-surface air velocity. Thus, the maximally buoyant limit should be considered primarily of theoretical interest rather than of practical value.
\end{remark}

To understand how aligment with Lagrangian motion is realized in the long run, we note that for sufficiently small inertial response time $\tau$, Eq.\@~\eqref{eq:MR} reduces time-asymptotically to
\begin{equation}
  \dot\x \sim \u + \tau \u_\tau,
  \label{eq:MRslow}
\end{equation}
where
\begin{equation}
  \u_\tau := R\frac{\D{\v}}{\D{t}} + R_\mathrm{a}\frac{\D{\va}}{\D{t}} + R \left(f + \tfrac{1}{3}\omega\right) \v^\perp + R_\mathrm{a} \left(f + \tfrac{1}{3}\omega_\mathrm{a}\right) \va^\perp- \frac{\D{\u}}{\D{t}} - \left(f + \tfrac{1}{3}R\omega + \tfrac{1}{3}R_\mathrm{a}\omega_\mathrm{a}\right) \u^\perp.
\end{equation}
This equation governs the motion on the \emph{slow manifold}, namely, the (2+1)-dimensional manifold $M_\tau := \{(\x,\vp,t) : \vp = \u(\x,t) + \tau\u_\tau(\x,t)\}$ embedded in the (4+1)-dimensional extended phase space $(\x,\vp,t)$. Rigorously speaking, $M_\tau$ represents a normally hyperbolic manifold that has the ability to exponentially attract all solutions of the complete Maxey--Riley equation,  Eq.\@~\eqref{eq:MR}.  A technical observation is that such an attraction cannot be expected to be monotonic when the carrying flow velocity $\mathbf u$ varies rapidly. To derive the reduce equation \eqref{eq:MRslow}, one must recognize that the complete equation, Eq.\@~\eqref{eq:MR}, comprises slow ($\x$) and fast ($\vp$) variables, enabling the application of geometric singular perturbation theory \cite{Fenichel-79, Jones-95}, extended to non-autonomous systems \cite{Haller-Sapsis-08}. This approach was utilized in \cite{Beron-etal-19-PoF, Beron-21-ND} for the BOM equation and can be straightforwardly applied on Eq.\@~\eqref{eq:MR} to obtain Eq.\@~\eqref{eq:MRslow}.

Note that when particles have neutral buoyancy, $\u_\tau \equiv 0$ and therefore Eq.\@~\eqref{eq:MRslow} reduces to $\dot\x \sim \v$.  In a similar fashion, for particles exhibiting maximum buoyancy, $\u_\tau \equiv 0$, reducing Eq.\@~\eqref{eq:MRslow} to $\dot\x \sim \va$. This elucidates the earlier mention of synchronization with Lagrangian motion over time.

\section{Final considerations}

We note that a Maxey--Riley theory can be developed using Eq.\@~\eqref{eq:MR} for systems of inertial particles interacting through a force, such as elastic coupling, which has been proposed for modeling \emph{Sargassum} seaweed transport \cite{Beron-Miron-20}. Various other types of interactions are also possible. For example, they may represent cohesion required to simulate the movement of oil or sediment particles. Regardless of the interaction type, the $i$th particle in a system of $N$ particles will evolve as described by Eq.\@~\eqref{eq:MR}, and in the long term by Eq.\@~\eqref{eq:MRslow}, with the coupling force
\begin{equation}
    \mathbf F_i = \sum_{j\in\operatorname{neighbor}(i)}\mathbf F(|\x_{ij}|),
\end{equation}
where $|\x_{ij}|$ is the distance between particles $i$ and $j$. The term $\mathbf F_i$ enters as an additional term in the right-hand-side of Eq.\@~\eqref{eq:MR} and of Eq.\@~\eqref{eq:MRslow}, but multiplied by $\tau$, under the implicit assumption that $\mathbf F_i = O(1)$ as measured by $\tau$.

Another remark involves the derivation of both the complete, Eq.\@~\eqref{eq:MR}, and simplified, Eq.\@~\eqref{eq:MRslow}, Maxey--Riley equations on a spherical surface. Consider the rescaled longitude--latitude coordinates $x = a_\odot\cos\vartheta_0(\lambda - \lambda_0)$ and $y = a_\odot(\vartheta - \vartheta_0)$, where $a_\odot$ is the Earth's average radius, and $(\lambda_0, \vartheta_0)$ is a reference point. We define $\gamma_\odot(y) := \sec\vartheta_0\cos\vartheta$ and $\tau_\odot(y) := a_\odot^{-1}\tan\vartheta$, which account for the geometric effects of Earth's spherical shape \cite{Ripa-JPO-97a}. The fluid vorticity 
\begin{equation}
  \omega_\mathrm{f} = \gamma_\odot^{-1}\partial_x v^y_\mathrm{f} - \partial_y v^x_\mathrm{f} + \tau_\odot v^x_\mathrm{f}. 
  \label{eq:vor-sph}
\end{equation} 
The covariant derivative of the fluid velocity $\vf$ along a fluid trajectory,
\begin{equation}
    \frac{\D{\vf}}{\d{t}} = \partial_t\vf + \gamma_\odot^{-1}v^x_\mathrm{f}\partial_x \vf + v^y_\mathrm{f}\partial_y \vf +  \tau_\odot v^x_\mathrm{f}\vf^\perp.
    \label{eq:dvdt-sph}
\end{equation}
To obtain the full Maxey--Riley equation on the sphere, one needs to:
\begin{enumerate}
    \item replace the vorticities in Cartesian coordinates in Eq.\@~\eqref{eq:MR}, given by Eq.\@~\eqref{eq:vor}, with those in spherical coordinates specified in Eq.\@~\eqref{eq:vor-sph}, and substitute the material derivatives in Eq.\@~\eqref{eq:dvdt} with covariant derivatives from Eq.\@~\eqref{eq:dvdt-sph}; and
    
    \item add the term $\tau_\odot v^x_\mathrm{p} \vp^\perp$ to the left-hand side of Eq.\@~\eqref{eq:MR}. 
\end{enumerate}
For deriving the reduced Maxey--Riley equation on the sphere, one should: 
\begin{enumerate}
    \item similarly replace the material derivatives in Eq.\@~\eqref{eq:MRslow} with covariant derivatives from Eq.\@~\eqref{eq:dvdt-sph}, extending $\vf$ to also represent $\u$; 
    
    \item subtract from the right-hand side of Eq.\@~\eqref{eq:MRslow} the term $\tau_\odot u^x \u^\perp$; and
    
    \item multiply the left-hand side of Eq.\@~\eqref{eq:MRslow} by $\smash{\sqrt{\mathsf{m}}}$, where the matrix $\mathsf{m}$ provides a representation of the spherical metric in the rescaled coordinates $(x,y)$ introduced above, with elements $\mathsf{m}_{11} = \gamma_\odot^2$, $\mathsf{m}_{11} = 1$, and $\mathsf{m}_{12} = 0 = \mathsf{m}_{21}$.
\end{enumerate}

\begin{remark}
  We have identified three inaccuracies in \textup{\cite{Beron-etal-19-PoF}} that we take the chance to rectify: 1) the initial equality in Eq.\@~(A5) should be omitted; 2) the term $\dot{x}$ in Eq.\@~(A8) must be scaled by $\sqrt{\mathsf m}$; and 3) the term $\tau_\odot v^1$ preceding $\frac{1}{3}R\omega$ in Eq.\@~(A8) should be replaced with $2\tau_\odot u^1$ (the discrepancy in the total derivative definition with respect that used in this paper results in the factor of 2). The first error was propagated to \textup{\cite{Beron-21-ND}}, as can be seen in Eq.\@~(39) in that paper.  The third error was also propagated to \textup{\cite{Miron-etal-20-GRL}}; however, in that study, the complete BOM equation was employed, as opposed to its reduced form applicable asymptotically on the slow manifold, thereby not affecting the results.
\end{remark}

\section*{Acknowledgments}

A recent collaboration with Jos\'e Barrientos and Luis Zavala has provided motivation to complete this work for publication, which has been done with support from the National Science Foundation (NSF) under grant number OCE2148499.

\section*{Author declarations}

\subsection*{Conflict of interest}

The author has no conflicts to disclose.

\subsection*{Author contributions}

This paper is authored by a single individual who entirely carried out the work.

\section*{Data availability}

Data sharing is not applicable to this article as no new data were created or analyzed in this study.

\bibliographystyle{alpha}
\newcommand{\etalchar}[1]{$^{#1}$}

\end{document}